\documentstyle[aps,prl,twocolumn,tighten,epsf]{revtex}

\begin{document}
\title{Voltage fluctuations on a superconductor grain attached to a
quantum wire}
%Boundary conformal field theory of low frequency
%equilibrium
%noise in a  normal-superconducting mesoscopic system}
\author{Masaki Oshikawa$^1$ and Alexandre M. Zagoskin$^2$}

\date{October 15, 1998}

\address{Department of Physics, Tokyo Institute of Technology$^1$,
Oh-oka-yama, Meguro, Tokyo 152-8511 JAPAN\\
Department of Physics and Astronomy, University of British Columbia$^2$,
Vancouver, BC V6T1Z1 CANADA}

\maketitle
\begin{abstract}
When a finite superconductor is in contact with a 1D normal conductor,
superconducting phase fluctuations lead
to power-law response of the normal
subsystem.
As a result, the charge fluctuations on the superconductor at zero
temperature have logarithmic correlator and a $1/\omega$ power spectrum
($1/f$-noise).
At higher temperatures $1/\omega$ is pushed to the high frequency
region, and $1/\omega^2$ behavior prevails.
\end{abstract}


% 74. Superconductivity
% 74.40.+k Fluctuations (noise, chaos, nonequilibrium superconductivity, localization, etc.)
% 74.80.Fp Point contacts; SN and SNS junctions
%
% 72. Electronic Transport in Condensed Matter 
% 72.70.+m Noise processes and phenomena
%
% 05. Statistical Physics and Thermodynamics
% 05.40.+j Fluctuation phenomena, random processes, and Brownian motion
%
% 11.10.Kk Field theories in dimensions other than four
% 11.25.Hf Conformal field theory, algebraic structures
%

Correlated character of transport in mesoscopic
systems strongly influences character of
current and charge fluctuations there. Examples are
non-Poissonian shot noise in quantum point contacts
and NS junctions(\cite{Beenakker} and references therein) and
giant current noise in superconducting quantum point
contacts\cite{MartinRodero,AverinImam,Dieleman}. Noise measurements may
provide information about quantum correlations unattainable by any other
means\cite{LesovikLevitov}.

In this paper we will show that equilibrium fluctuations in a
mesoscopic NS system reveal the many-body reaction of
a conducting system to a sudden external perturbation, thus linking
the noise to such phenomena as Fermi edge singularity (FES)\cite{Mahan}
and Anderson orthogonality catastrophe\cite{Anderson}. Unlike the
standard situation when the perturbation is due to creation of
a scatterer, here the normal subsystem reacts to the change in the
{\em off-diagonal} pairing potential in the superconductor, to which
it is coupled via Andreev reflection processes on the NS
boundary\cite{Andreev,ZA}.

More specifically, consider the system shown in Fig.1. It consists of
a mesoscopic superconducting grain $S$ in contact with a 1D normal wire $N$.
We assume that the grain contains a large number of Cooper pairs,
$\left<n\right> \gg 1$, and its capacitance $C$ is large enough
(so we are in the opposite limit to the Coulomb blockade
regime\cite{Tinkham}), but the contact region is smaller
than the Josephson screening length $\lambda_J$, so that
both the modulus and phase of the order parameter
$\Delta = |\Delta|e^{i\phi}$ can be taken constant there.
Temperature is low enough, $T\ll|\Delta|$, to neglect quasiparticle
excitations in the superconductor.
Finally, we assume that the 1D conductor is scattering free,
and its length $L \gg \xi_0$, where $\xi_0$ is  the superconducting
coherence length.

If isolated, the superconducting phase in the grain would fluctuate,
the system being in the state with fixed particle number (see e.g.
\cite{Tinkham}, Ch.7). To the 1D conductor brought in contact with
the grain, these phase fluctuations will translate into fluctuating
Andreev boundary condition\cite{Andreev}.

First let us present a heuristic picture of the phenomenon at zero
temperature.
The effect of a sudden change in a boundary condition is a slow,
power-law relaxation of the normal system to its new ground state, that
is, Fermi edge singularity. It  is conveniently described
in terms of boundary conformal field theory, which allows a
unified description of FES and Anderson orthogonality
catastrophe\cite{AL}.
Both are characterized by scaling dimension $x$ of the boundary
condition changing
operator $\cal O$, so that the correlation function
$\left<{\cal O}^{\dagger}(\tau){\cal O}(0)\right> \sim  \tau^{-2x},$
and the dimension $x$ itself is related to the finite-size ($O(1/L)$)
correction to the ground state energy shift due to the action
of $\cal O$:
\begin{equation}
x = \frac{L}{\pi} \Delta E_{1/L}.
\label{dimX}
\end{equation}
Hereafter we set $\hbar=1$ and the velocity of the excitations
(Fermi velocity in the non-interacting case)
$v=1$ unless stated otherwise; $L$ is the
characteristic size of the system (wire length in our
case). This relation holds both for usual potential scattering
and Andreev scattering\cite{ZA}.

The boundary condition changing operator
in our case  changes the  phase of superconductor by $\Phi$:
$
{\cal O}_{\Phi} = \exp{[\Phi \frac{\partial}{\partial\phi}]} =
\exp{[i\Phi \hat{n}]}.
$
Here we used the relation between $\phi$
and  the Cooper pair number operator on the grain,
$\hat{n} = \frac{1}{i}\frac{\partial}{\partial\phi}$, which holds if
$\left<n\right> \gg 1$\cite{Tinkham}.
In agreement with general CFT argument, the Green's function of
${\cal O}_{\Phi}$ is
\begin{equation}
G_{\Phi}(\tau) = \left<{\cal O}_{\Phi}^{\dagger}(\tau){\cal
O}_{\Phi}(0)\right> \equiv
\left<e^{-i\Phi\hat{n}(\tau)}e^{i\Phi\hat{n}(0)}\right>
= (\tau/\tau_0)^{-2x}, \label{GRIN}
\end{equation}
with some cutoff parameter $\tau_0$. Though by itself $G_{\Phi}$ has no
direct
physical meaning, it is related to the correlator of charge
fluctuations on the grain:
$
K_Q(\tau)/(4e^2) = \left<
 \left[\delta\hat{n}(\tau),\delta\hat{n}(0)\right]_+
   \right>/2 ,
$
where $\delta\hat{n} = \hat{n}-\left<\hat{n}\right>.$
Obviously,
\begin{eqnarray}                    \frac{K_Q(\tau) - K_Q(0)}{4e^2}
\equiv
\frac{1}{2}\left<\left[\delta\hat{n}(\tau),\delta\hat{n}(0)\right]_+\right>
-
\left<\delta\hat{n}^2\right> =\nonumber \\
\frac{1}{2}\left<\left[\hat{n}(\tau),\hat{n}(0)\right]_+\right> -
\left<\hat{n}^2\right> =
\lim_{\Phi\to 0}\frac{\frac{1}{2}
\left(G_{\Phi}(\tau)+G_{\Phi}(\tau)^*\right) -1}{\Phi^2}.\label{CORR}
\end{eqnarray}
Charge fluctuations are measured by e.g. monitoring the electrostatic
potential of the grain.

The dimension $x$ is easily found\cite{ZA} if
notice that the energy correction entering (\ref{dimX}) is simply the
Josephson energy of a ballistic 1D SNS junction of length $L$,
with phase difference $\Phi$\cite{Ishii,SAB} at zero temperature:
$ E_J(\Phi) =  \Phi^2 / (4 \pi L)$,
so that
$ x = L E_J(\Phi) / \pi  = \left[ \Phi/(2\pi) \right]^2$.
Substituting this and (\ref{GRIN}) into (\ref{CORR}), we find for the
charge fluctuations on the grain
\begin{equation}
\frac{K_Q(\tau)}{4e^2}  =
\left<\delta\hat{n}^2\right> - \frac{1}{2\pi^2}\ln\frac{\tau}{\tau_0}.
\label{LOG}
\end{equation}
For the spectral density of charge fluctuations,
$S_Q(\omega)=2\int_0^{\infty} K_Q(\tau)\cos(\omega\tau) d\tau,$
such logarithmic behavior of correlator
translates into $S_Q(\omega) \propto 1/\omega$ ($1/f$-noise) in
the corresponding frequency interval; it
is often observed in spin glasses, magnetic polycrystals and other
systems with wide distribution of relaxation times\cite{1/f,Dutta}.
In our case the source
of slow relaxation times is the normal system with its power-law
approach to the new ground state.
The idea of invoking  infrared singularity to explain  scaling that
underlies current $1/f$-noise observed  in normal conductors
was exploited in e.g. model of
``quantum 1/f-noise" due to  Brem\ss trahlung, but it
encountered fatal difficulties (see \cite{Dutta} and references
therein). In contrast, we see no fundamental difficulty in the
present mechanism, which will be shown to be consistent with
fluctuation-dissipation theorem.

Expression (\ref{LOG}) is obviously only an intermediate asymptotics
for $1 \ll \tau/\tau_0 \ll \exp[2\pi^2\left<\delta\hat{n}^2\right>].$
Therefore we need a more consistent approach, taking into account
effects of charging energy and finite temperature.
We will see
that the region of $1/f$-fluctuations is pushed by thermal effects
to high frequencies, and $1/\omega^2$-type
spectrum prevails except at very low temperature.

We will describe
the wire in terms of Tomonaga-Luttinger liquid (TLL), which allows
us to treat simultaneously non-interacting (1D Fermi-liquid) and
interacting case of such S-TLL system\cite{fazio}. 
Though as we will see interactions do not change qualitative picture
of the phenomenon, they may lead to interesting side effects.

Spin and charge degrees of freedom in TLL separate.
Since spin is totally reflected by NS boundary both in the process of ideal
Andreev reflection and ideal normal reflection, it is irrelevant
and will be dropped.
For the charge, the Lagrangian density is
$ {\cal L} = 1/(2 \pi K)
[ (\partial_t \theta)^2 - ( \partial_x \theta)^2 ] ,
$
where $K$ is the coupling constant, which is unity
for non-interacting electrons.

Physically, the boson field $\theta(x,t)$ corresponds to the phase
of Charge Density Wave (CDW),
while its dual $\phi(x,t)$   to the superconducting phase of the
Cooper pair\cite{Mahan}.
For a perfect NS interface, the boundary value of the superconducting
phase of the TLL should be equal to that of the superconductor,
which in the absence of phase fluctuations gives a Neumann boundary
condition for the   field  $\theta(x,t)$ in the wire~\cite{OZ},
equivalent to a Dirichlet boundary
condition on $\phi$: $\phi(0,t) = \Phi = \mbox{const}$.

The charge $Q$ on the grain
can be related to the boundary value of the ``CDW''
field $\theta$.
Indeed, the current in  TLL can be written as
\begin{equation}
 J(x,t) =  - \frac{\sqrt{2} e}{\pi}
  \frac{\partial \theta(x,t)}{\partial t}.
\label{eq:current}
\end{equation}
Therefore the electric charge of the superconductor
at time $t$ is given by
\begin{equation}
 Q(t) = Q(-\infty) - \int_{-\infty}^t J(0,t) dt
  = \frac{\sqrt{2} e}{\pi } \theta(0,t),
\end{equation}
where we have defined $Q$ and $\theta$ so that
$Q=\theta(0,t)=0$ corresponds to the neutral superconductor grain.
Similar identification was used in studies of
Coulomb Blockade~\cite{Matveev},
although our application is to quite different regime.

Including the charging energy associated with $Q$,
$ E_Q  = \frac{Q^2}{2C} = \frac{e^2}{\pi C^2} \theta(0,t)^2 $,
we arrive at the effective action of the problem:
\begin{equation}
 S = \int \int dt dx  \frac{1}{2} ( \partial_{\mu} \Theta )^2
  - \int dt \frac{u}{2} \Theta(0,t)^2.
\label{eq:effaction}
\end{equation}
Here   $\Theta = \theta / \sqrt{\pi K}$
and $u = 2 e^2 K/ (\pi C)$
($u = 2 e^2 K/ (\pi\hbar C)$ in conventional units).
This is a free boson action with a mass term only on the boundary,
and this effective theory is exactly solvable.
The calculation of the charge fluctuation in the superconductor
is reduced to the calculation of the correlation function of the
boundary field $\Theta(0,t)$. The physics of the problem is
contained in a proper choice of the boundary condition at $x=0$
(on the NS boundary)\cite{multichannel}.

The Matsubara Green's function of $\Theta$ (for $\tau = i t$)
$
 {\cal G}(x,\tau) = -\langle T_{\tau} \Theta(x,\tau) \Theta(0,0) \rangle ,
$
is  the Green's function of the Laplacian on the plane,
with the   boundary conditions
$ \partial_x \Theta(x=0,\tau) = u \Theta(x=0,\tau)$
and
$\Theta(x,\tau + \beta) = \Theta(x,\tau)$, where
$\beta = 1/T$ is the inverse temperature.
As a consequence, the Green's function $\cal G$ for the source
at $(x_0,0)$ satisfies
\begin{eqnarray}
( {\partial_x}^2 + {\partial_{\tau}}^2  )
    {\cal G}(x,\tau) &=&  \delta(x-x_0) \delta(\tau) \label{y00}
\\
\partial_x {\cal G}(x=0,\tau) &=& u{\cal G} (x=0,\tau) .\label{x00}
\end{eqnarray}
Because we are interested in the correlation function of the boundary,
we take the limit $x_0 \rightarrow 0$. Integrating (\ref{y00})
over $x$ from $0$ to $x_0+0$, we see that the boundary condition
(\ref{x00}) must be now modified:
\begin{equation}
\partial_x {\cal G}(x=0,\tau) = u{\cal G}(x=0,\tau) - \delta(\tau).
\label{x000}
\end{equation}
For Fourier components we obtain an ordinary
differential equation:
$\partial_x^2 {\cal G}(x,\omega_n) = {\omega_n}^2 {\cal G}$
with the boundary condition
$\partial_x {\cal G}(x=0,\omega_n) = u {\cal G}(x=0,\omega_n) - 1.$
Imposing the asymptotic condition
$\lim_{x \rightarrow \infty} {\cal G}(x,\tau) \rightarrow 0 $,
we find
\begin{equation}
 {\cal G}(x,\omega_n) =
  \frac{-1}{u + | \omega_n |} e^{-|\omega_n| x}.
\label{eq:matsubara}
\end{equation}
The Matsubara Green's function for the boundary field is
${\cal G}(x=0,\omega_n) = - 1/ (u + | \omega_n|)$.

The correlation functions in the real time are obtained by an analytic
continuation from the Matsubara Green's function.
In a standard way, we obtain the retarded and advanced
Green's functions, $ G^{R(A)} (\omega) = -1/(u \mp i \omega)$.
Their mismatch at the real axis, which is pure imaginary, gives the
spectral density of the system:
$\rho(\omega) = - 2 {\rm Im} G^R(\omega) = 2 \omega/(u^2 + \omega^2)$.
The spectrum of electrostatic potential fluctuations
on the grain  is given by:
\begin{eqnarray}
\lefteqn{S_V(\omega) = \frac{S_Q(\omega)}{C^2} =
\int_{-\infty}^{\infty}
 \frac{1}{2 C^2} \langle Q(t) Q(0) + Q(0)Q(t) \rangle e^{i \omega t} dt}
 \nonumber \\
&&
= \frac{2 K e^2}{\pi C^2} \rho(\omega)
\frac{1}{2} \coth{(\frac{ \beta \omega}{2})}
=
\frac{2 K e^2}{\pi C^2} \frac{\omega}{u^2 + \omega^2}
\coth{( \frac{\beta \omega}{2})}.
\label{FDT}
\end{eqnarray}
This result has the form expected from fluctuation-dissipation theorem
(FDT)\cite{LifshitsIX}.

In the low temperature/high frequency limit $\beta \omega    \gg 1$,
the spectrum is proportional to $ |\omega|/(u^2 + \omega^2)$.
At frequency  much higher than the ``infrared cutoff" $u$,
the spectrum is
indeed $1/\omega$ as expected from the heuristic argument~(\ref{LOG}).
On the other hand, at smaller $\omega$  fluctuations are suppressed,
because  finite capacitance obviously excludes  large fluctuation
of the charge in the grain.
The $1/\omega$ spectrum, which is independent of the interaction
parameter $K$. It automatically arises from the free boson
field theory once the boson field itself is identified with a physical
observable, because the correlation function of the
boson field is logarithmic.
We have thus shown an example of ``quantum $1/f$ noise''
naturally derived within the framework of equilibrium statistical
mechanics and field theory.
Although the $1/\omega$ behavior is only expected at
very low temperature and high frequency,
we believe this presents certain interest both from the point of view
of translation of infrared divergences into $1/f$-spectrum, and of
its destruction by thermal fluctuations.

At higher temperature/lower frequency $\beta \omega \gg 1$, the
spectrum is proportional to $ T / (u^2 + \omega^2)$.
The spectrum is (restoring the fundamental constants)
$4 kT K e^2 / (\pi  \hbar \omega^2)$
for frequency larger than inverse capacitance $\omega \gg  u$.
In the opposite limit of small frequency $\omega \ll u$,
(but still $ \omega \beta \gg 1$),
the fluctuation is simply a white noise with the magnitude
$ 2 R_H kT/(4K)$ where $R_H = h/e^2$, again in a complete agreement with FDT.
The transition from   low- to high- temperature behavior
occurs at   $\beta \omega \sim 1$, namely $k T \sim hf$ (see Fig.\ref{figZ}).
For $T=10 \mbox{mK}$, the boundary is at $f \sim 200 \mbox{MHz}$.
This would be within the capabilities of modern experiment;
we expect the crossover between $1/\omega^2$ and $1/\omega$ behavior
could be directly observable.

The charge fluctuation in the grain means the current flowing through
the wire also fluctuates.
The correlation function of the current can be obtained by
a similar method. For example, the Matsubara Green's function of
of the current at $J(x_0)$ is given by
$ ( 2 e^2 \omega_n^2 / \pi^2 ) {\cal G}_{x_0}(x_0,\omega_n)$,
where
\begin{equation}
{\cal G}_{x_0}(x_0, \omega_n) =
- \frac{1}{2 |\omega_n|}
\left[
 1 - \frac{u-|\omega_n|}{u+|\omega_n|} e^{-2 |\omega_n| x_0}
\right]
\end{equation}
is the Green's function of the Laplacian with the source at $x_0$.
The $1/\omega$ fluctuation of the grain voltage translates to
an $\omega$-linear fluctuation of the current, although a
similar behavior persists even for the current far from the grain.

So far, we have considered a semi-infinite wire connected to the
grain.  In the remainder of this paper, we discuss how our results are
affected by wire being finite.

Instead of manufacturing a very long wire, it would be easier to
connect the other end of wire to a normal reservoir.
We assume that the junction to the reservoir is adiabatic (no
backscattering), and the phase information is immediately lost in the
reservoir (perfect thermalization). Then the situation
is not qualitatively different from the case of the semi-infinite wire,
since the particles can leave to infinity (reservoir), and
the usual reasoning of Landauer formalism applies\cite{LANDAUER};
the transport of the charge from/to the grain controlled by
the geometric restriction (narrow 1D channel) implies that our theory
based on 1D holds, even though the system is not purely 1D.
Thus we expect our results for the semi-infinite wire applies
to the more realistic case of the finite wire connected to
a normal reservoir, if the interaction effects in the wire
can be neglected.

When the interaction in the wire is not negligible, the problem is
more subtle, since a quasiparticle excitation inside the wire
is a collective motion of many electrons and cannot be simply
transferred to the reservoir.
While a rigorous theory of junction between a TLL and
normal reservoir is not yet established, it has been studied by
modelling the reservoir with a (semi-infinite) 1D Fermi liquid.
As a simplest model, we study a finite wire of length $L$, in which
the coupling constant is $K$, attached to semi-infinite wire
with coupling constant $K'$, $K \equiv \alpha K'$.
To model the reservoir by a 1D Fermi liquid, $K'$ should be unity
but we shall obtain the solution for general values of $K'$
because it can be applied also to other interesting cases.
It has been found that normal and Andreev-type reflections
occur at the boundary if $K \neq K'$~\cite{Safi}.

By a similar line of arguments as before,
we obtain the Matsubara Green's function of the boundary field $\Theta$ as
\begin{eqnarray}
\lefteqn{{\cal G}(0,\omega_n) =} \nonumber\\
&& \frac{- [\cosh{(\omega_n L)} + \alpha \sinh{(|\omega_n| L)}]}{
 (u + \alpha | \omega_n |) \cosh{(\omega_n L)} +
 ( |\omega_n| + \alpha u) \sinh{(\omega_n L)}
     }
\label{eq:calGlen}
\end{eqnarray}
Since the temperature effect is given by the same FDT factor
$\coth{(\beta \omega/2)}/2$ as in the semi-infinite wire case,
we concentrate on how the spectral
density $\rho(\omega)=-2{\rm Im}\: G^R(\omega)$ is affected.

For $\alpha=1$, the result reduces to the semi-infinite wire case,
as it should be.
For $\alpha \approx 1$, the overall shape of the
spectral density remains the same.
However, a variation with the period $1/L$ is superimposed on the
dependence for semi-infinite wire (Fig.\ref{figQ}).
This is the resonant structure due to (normal and Andreev)
reflections at the boundary between two wires.
If the junction between the wire and the reservoir is smooth and not
abrupt, we expect that the resonant structure is smeared out;
then the observed spectrum would be quite close to the
prediction for the semi-infinite wire.

If $\alpha$ far from $1$, the resonant structure dominates.
In particular, the case $K' \rightarrow 0$ (or $K' \rightarrow \infty$)
is equivalent to a finite wire with open end (or a finite wire attached to
a very large superconductor reservoir), respectively.
In these limits $\alpha \rightarrow \infty$,
($\alpha \rightarrow 0$) the spectral density develops into a set of
discrete $\delta$-functions  with peak spacings of order $1/L$,
rather than a continuous function of $\omega$.
This reflects the finiteness  of the system.
In the $L \rightarrow \infty$ limit the $\delta$-functions become dense
and converge to the smooth function we have obtained for the
semi-infinite wire.

It is also instructive to take the opposite limit of short wire
$L \rightarrow 0$.
For the open end case ($K' \rightarrow 0$), the resonances
are found at $\omega \sim (2n - 1 ) \pi/(2L)$ where $n$ is a
natural number.
For the wire connected to the superconductor reservoir,
in the short-wire limit,
the resonances are found at $\omega \sim n \pi/L$, $n=1,2,\dots$.
The resonance which would correspond to $n=0$ is actually affected by
the finite capacitance, and its location is  given by
$\omega = 1/\sqrt{LC}$.
This can be simply understood from the following argument: from
the conjugate relation between the phase and the charge,
the effective Hamiltonian of the system is given by
(see e.g. \cite{Tinkham})
$
 \hat{H} = - \frac{1}{2C} (\frac{\partial}{\partial \Phi})^2
  + \frac{1}{2L} (\Phi - \Phi_0)^2 ,
$
where $\Phi$ is the superconducting phase of the grain,
the first and second term represent the charging energy of the grain
and the Josephson energy of the wire, respectively.
This is a Hamiltonian of a harmonic oscillator with the eigenfrequency
$1/\sqrt{LC}$, exactly where the resonance is found.

In conclusion, we have shown that quantum phase
fluctuations in a superconducting grain in contact with a normal wire
are translated into equilibrium charge fluctuations in the system,
with specific spectrum demonstrating transition from $1/\omega^2$
to $1/\omega$ behavior at very low temperature.
The mechanism of fluctuations is reaction of the normal subsystem
to sudden change of the {\em off-diagonal} pairing
potential  on the NS boundary (Fermi edge singularity),
leading in the low-temperature limit to quantum $1/f$ noise.
Similar effect can be expected to exist also in systems with
superconducting grain in contact with 2D conductors under
appropriate conditions.

While the experimental observation of this effect presents a certain
challenge, we hope it to be possible in near future.
Concerning the experiments, impurity scatterings and
physical effects of probing will require further theoretical study,
which is however beyond scope of the present paper.

We are grateful to I. Affleck, N. Kokubo and S. Okuma for valuable
discussions and to M. B\"{u}ttiker and R. Fazio for helpful comments on the
manuscript.

\begin{figure}

\epsfxsize=3 in
\epsfbox{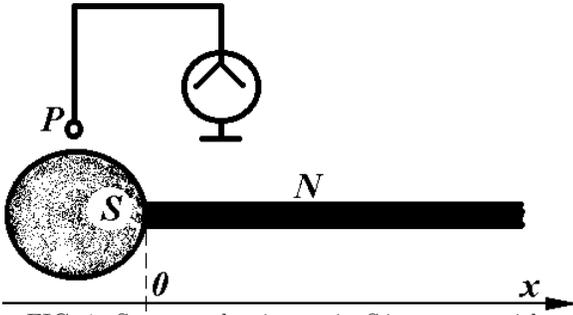}
\caption{Superconducting grain $S$ in contact with a 1D conducting
wire $N$.
Tip $P$
probes the electrostatic potential of the grain.}

\label{figY}
\end{figure}

\begin{figure}

\epsfxsize=3 in
\epsfbox{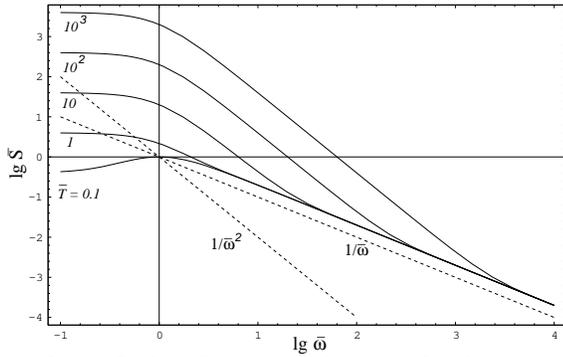}
\caption{Reduced spectral density of voltage noise in a grain, 
$\bar{S}(\omega) = S_V(\omega)/[Ke^2/\pi c^2]$ (Eq.(12)), as a function of
$\bar{\omega} \equiv \omega/u$ for different values of $\bar{T} \equiv T/u$.}

\label{figZ}
\end{figure}

\begin{figure}

\epsfxsize=3 in
\epsfbox{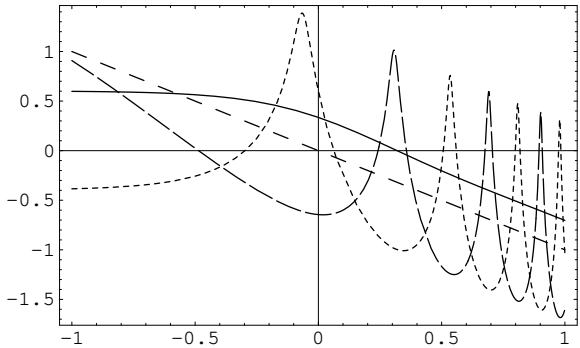}
\caption{ Resonant structure of spectral density for $\alpha = 0.1$ (dots),
$\alpha = 1$ (solid line), and $\alpha = 10$. 
Dimensionless length of the wire $L = 1;\:\: \bar{T}=1$. Dashed straight line
illustrates $1/\bar{\omega}$ dependence.}

\label{figQ}
\end{figure}

\end{document}